\documentclass{jaa}
\usepackage{natbib}
\usepackage{graphicx}
\usepackage{color}

\newcommand{\la}{\left\langle}
\newcommand{\ra}{\right\rangle}
\newcommand{\be}{\begin{equation}}
\newcommand{\ee}{\end{equation}}
\newcommand{\bse}{\begin{subequations}}
\newcommand{\ese}{\end{subequations}}
\newcommand{\bea}{\begin{eqnarray}}
\newcommand{\eea}{\end{eqnarray}}
\newcommand{\ba}{\begin{array}}
\newcommand{\ea}{\begin{array}}

\begin{document}\sloppy

\title{Magnetohydrodynamic Turbulence: Chandrasekhar’s Contributions \& Beyond }


\author{Mahendra K. Verma\textsuperscript{1}}
\affilOne{\textsuperscript{1}Department of Physics, Indian Institute of Technology Kanpur, Kanpur 208016, India.\\}


\twocolumn[{

\maketitle

\corres{mkv@iitk.ac.in}


\begin{abstract}
In the period of 1948-1955, Chandrasekhar wrote four papers on magnetohydrodynamic (MHD) turbulence, which  are the first set of papers in the area. The field moved on following these pioneering efforts. In this paper, I will briefly describe important works of MHD turbulence, starting from those by Chandrasekhar.
\end{abstract}


}]


\doinum{12.3456/s78910-011-012-3}
\artcitid{\#\#\#\#}
\volnum{000}
\year{0000}
\pgrange{1--}
\setcounter{page}{1}
\lp{1}

\section{Introduction}

Chandrasekhar pioneered  the following areas of astrophysics: white dwarfs, neutron stars, black holes, stellar structures, radiative transfers, random processes, stability of ellipsoidal figures of equilibrium,  instabilities, and turbulence. His work on turbulence is not as well known as others, even though his papers on quantification of structure functions and energy spectrum of magnetohydrodynamic (MHD) turbulence are first ones in the field. Recently, \citet{Sreenivasan:ARFM2019} wrote a very interesting review  on Chandrasekhar's  contributions to fluid mechanics, which includes hydrodynamic instabilities and turbulence.  While  Sreenivasan's article is focussed on hydrodynamics,  in this paper, I will provide a brief review on Chandrasekhar's work in MHD turbulence.

In the years 1948 to 1960, Chandrasekhar worked intensely on turbulence.  In 1954, Chandrasekhar gave a set of lectures on turbulence in Yerkes Observatory.  These lectures, published  by \citet{Spiegel:book_edited:Turbulence},  illustrate Chandrasekhar's line of approach to understand turbulence.  To quote \citet{Spiegel:book_edited:Turbulence}, ``Still, Chandra pulled things together and published two papers on his approach (in 1955 and 1956). The initial reception of the theory was positive. Indeed, Stanley Corrsin once told me that, back in the mid-fifties, he was so sure that the `turbulence problem' would soon be solved that he bet George Uhlenbeck five dollars that he was right. Afterwards, when Corrsin and Uhlenbeck heard Chandra lecture on his theory, Uhlenbeck came over and handed Corrsin a fiver. It soon appeared that Uhlenbeck should have waited before parting with his money." Refer to \cite{Sreenivasan:ARFM2019} for a more detailed account of Chandrasekhar's work on turbulence and instabilities. The books by \cite{Wali:book:Chandra_bio} and \cite{Miller:book:Chandra} are excellent biographies of Chandrasekhar.

In this paper,  I  provide a brief overview of the leading  works in MHD turbulence, starting from those of Chandrasekhar.  These works are related to the inertial range of homogeneous MHD turbulence.  The works beyond Chandrasekhar's contributions are divided in two periods: (a) 1965-1990, during which the field was essentially dominated by the belief that Kraichnan-Iroshnikov model works for MHD turbulence.   (b) 1991-2010, during which many models and theories came up that support Kolmogorov-like spectrum for MHD turbulence. I also remark that the present paper is my personal perspective that may differ from those of others.

The outline of the paper is as follows: In Section 2, I will briefly introduce the theories of hydrodynamic turbulence by Kolmogorov and Heisenberg. 
 Section 3 contains a brief summary of Chandrasekhar's work on MHD turbulence that occurred between 1948 and 1955.  In Sections 4 and 5, I will brief works on MHD turbulence during the periods 1965-1990 and 1991-2010 respectively. Section 6 contains a short discussion on possible  approaches for resolving the present impasse in MHD turbulence. 
  I conclude in Section 7.

\section{Leading turbulence models before Chandrasekhar's work}
\label{sec:HD_turb}
Chandrasekhar worked on hydrodynamic (HD) and magnetohydrodynamics (MHD) turbulence  during the years 1948 to 1960.  Some of the papers written by him during this period are~\citet{Chandrasekhar:PTRS1950,Chandrasekhar:PRSA1951,Chandrasekhar:PRSA1951_II,Chandrasekhar:PTRS1952,Chandrasekhar:PRSA1955_I,Chandrasekhar:PRSA1955_II,Chandrasekhar:PR1956}. In addition, \citet{Chandrasekhar:book:Instability} wrote a famous treatise on hydrodynamic and magnetohydrodynamic instabilities. Chandrasekhar's lectures on turbulence  (delivered in 1954) have been published by \citet{Spiegel:book_edited:Turbulence}.
Refer to \cite{Sreenivasan:ARFM2019} for commentary on these works.

 During or before Chandrasekhar's work, there were important results by Taylor, Batchelor, Kolmogorov,  Heisenberg, among others.  Here, we briefly describe the turbulence theories of Kolmogorov and Heisenberg, primarily because Chandrasekhar's works on MHD turbulence are related to these theories.  We start with Kolmogorov's theory of turbulence.

\subsection{Kolmogorov's theory of turbulence}
\label{sec:Kolm}

Starting from Navier-Stokes equation,  under the assumptions of homogeneity  and isotropy, \citet{Karman:PRSA1938} [also see  \cite{Monin:book:v2}] derived  the following evolution equation for $\la u_i u'_i \ra$:
\bea
\frac{\partial}{\partial t} \frac{1}{2}  \la u_i u_i' \ra & = &
 \frac{1}{4} \nabla_l \cdot \la |{\bf u'-u}|^2 ({\bf u'-u}) \ra  +   \la F_{\mathrm{LS},i}  u_i'  \ra  \nonumber \\
 && + \nu \nabla^2 \la u_i u_i' \ra \nonumber \\
& = & T_u({\bf l}) + \mathcal{F}_\mathrm{LS}({\bf l}) - D_u({\bf l}),
\label{eq:Karman}
\eea    
where \textbf{u} and \textbf{u'} are the velocities at the locations \textbf{r} and \textbf{r+l} respectively, and $ \nu $ is the kinematic viscosity (see Figure 1). The terms $ T_u({\bf l})  $ and $ D_u({\bf l}) $ represent respectively the nonlinear energy transfer and the dissipation rates at scale \textbf{l}, while $  \mathcal{F}_\mathrm{LS}({\bf l}) $ is the energy injection rate by the external force $ \mathbf{F}_\mathrm{LS} $, which is active at  large scales. For a steady turbulence, under the limit $ \nu \rightarrow 0$, \citet{Kolmogorov:DANS1941Structure,Kolmogorov:DANS1941Dissipation} showed that in the inertial range (intermediate scales between the forcing and dissipation scales),
\bea
 \la [{\bf (u'-u) } \cdot \hat{\bf l}]^3 \ra  = -\frac{4}{5} \epsilon_u l, 
 \label{eq:K41}
\eea
where $ \epsilon_u $ is the viscous dissipation rate per unit mass, and $  \hat{\bf l} $ is the unit vector along \textbf{l}. Kolmogorov's theory is commonly referred to as \textit{K41} theory.  

\begin{figure}
	\centering\includegraphics[scale=1]{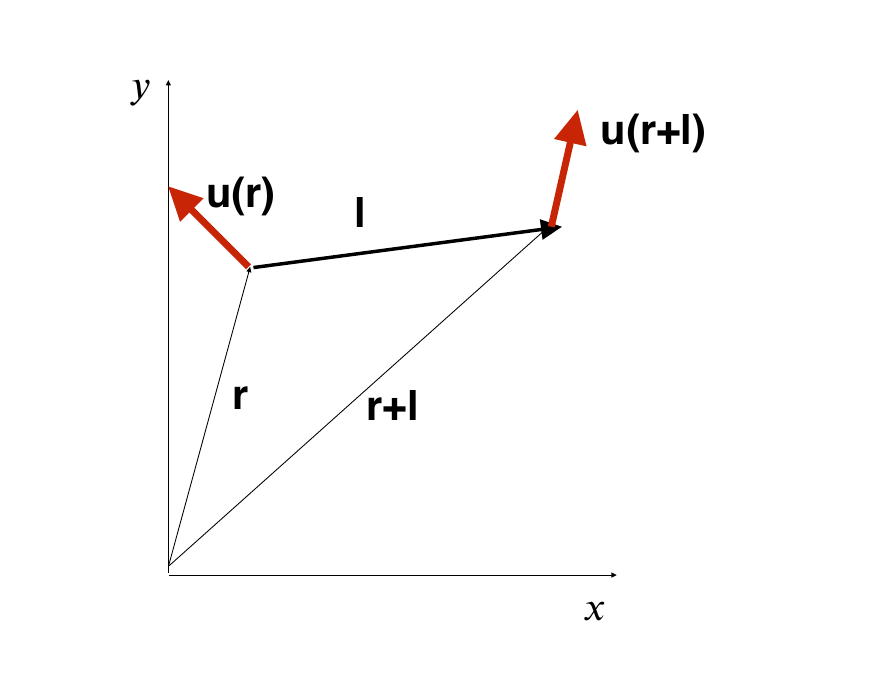}
	\caption{ The velocity fields at two points \textbf{r} and \textbf{r+l} are \textbf{u(r)} and \textbf{u(r+l)} respectively. We denote them using \textbf{u} and \textbf{u'} respectively.
	}
\end{figure}
A simple-minded extrapolation of  Eq.~(\ref{eq:K41}) leads to 
\bea
\la [{\bf (u'-u) } \cdot \hat{\bf l}]^2 \ra \approx  \epsilon_u^{2/3} l^{2/3} 
\eea
whose Fourier transform leads to the following  formula for the energy spectrum:
\bea
E(k) = K_\mathrm{Ko} \epsilon_u^{2/3} k^{-5/3},
\eea
where $ K_\mathrm{Ko} $ is a nondimensional constant [\cite{Frisch:book}].

In Fourier space, Eq.~(\ref{eq:Karman})  transforms to  the following energy transfer relation~(\cite{Verma:book:ET,Verma:JPA2022}):
\bea
\frac{\partial}{\partial t} E_u({\bf k},t) = T_u({\bf k},t) +  \mathcal{F}_\mathrm{LS}({\bf k},t) - D_u({\bf k},t),
\label{eq:energy_k_dot}
\eea
where $ E_u({\bf k}) = |{\bf u(k)}|^2/2$ is the modal energy, and
\bea
T_u({\bf k},t)  = \sum_{\bf p} \Im[ \{ {\bf k \cdot u(q)}\}  \{ {\bf u(p) \cdot u^*(k)}\} ],
\label{eq:Tk}
\eea
with $ {\bf q = k-p} $, represents the total energy gained by $ {\bf u(k)} $ via nonlinear energy transfers.  For isotropic turbulence, 
\bea
T_u({\bf k}) = T_u(k) = - \frac{d}{dk} \Pi_u(k),
\eea
where $ \Pi_u(k) $ is the energy flux emanating from a wavenumber sphere of radius $ k $. In the inertial range,  the energy injection by the external force vanishes and viscous dissipation rate is negligible, hence $ \Pi_u(k) \approx \mathrm{const} $. Refer to \cite{Frisch:book},  \cite{Verma:book:ET}, and \citet{Verma:JPA2022} for more details.

\subsection{Heisenberg's theory of turbulence}
In this subsection, we describe Heisenberg's theory of turbulence because Chandrasekhar employed this theory to derive the energy spectrum for MHD turbulence.  \cite{Heisenberg:PRSA1948} derived an integral equation for the temporal evolution of kinetic energy spectrum $ E_u(k) $ under the assumption of homogeneity and isotropy. In particular, he derived that
\bea
\frac{\partial}{\partial t}  \int_0^k dk' E_u(k',t) & = & -2 \left(\nu+\alpha \int_k^\infty \sqrt{\frac{E_u(k')}{k'^3}} dk' \right) \nonumber \\
&& \times \int_0^k k'^2 E_u(k') dk'.
\label{eq:Heisenberg}
\eea
In the above equation, the second term of the right-hand-side is a model for  the diffusion of kinetic energy to smaller scales by eddy viscosity (induced by the nonlinear term).  Many authors, including Chandrasekhar, have employed Heisenberg's model  for modelling turbulent flows.

\section{Chandrasekhar's contributions to MHD turbulence}
\label{sec:Chandra}
Chandrasekhar wrote around a dozen  papers on  turbulence, four of which are on MHD. He focussed on the closing the hierarchical equations of turbulence. In the following, I will provide a brief overview of Chandrasekhar's work on MHD turbulence.

In  turbulence, the nonlinear interactions induce energy transfers among the Fourier modes.  Hydrodynamic interactions involve triadic interactions, e.g., 	in Eqs.~(\ref{eq:energy_k_dot},\ref{eq:Tk}), the Fourier mode $ {\bf u(k)} $ receives energy from the Fourier modes $ {\bf u(p)} $ and $ {\bf u(q)} $. 
 In 1954, Chandrasekhar gave a set of lectures on turbulence in which he showed that the energy transfer from $ {\bf u(p)} $ to $ {\bf u(k)} $ with the mediation of  $ {\bf u(q)} $ is
	\bea
	Q({ \bf k, k'} ) = \Im [({\bf u({q})  \cdot k } ) ({\bf u{(p} ) \cdot u^*({k}) })].
	\label{eq:Qkp}
	\eea
As far as we know, the above formula first appears in~\cite{Onsagar:Nouvo1949_SH}, but not in any paper of Chandrasekhar.  Incidently,  Onsager is not cited for this formula in \citet{Spiegel:book_edited:Turbulence}. Hence, it is not apparent whether Chandrasekhar derived Eq.~(\ref{eq:Qkp}) independently, or he was aware of Onsager's work.  Around 2000, we were working on the energy fluxes of MHD turbulence, and we~(\cite{Dar:PD2001}) arrived at the same formula independently. Note that  in MHD turbulence, energy transfers occur between velocity field and magnetic field as well.

Let us get back to MHD turbulence. Chandrasekhar's four papers on MHD turbulence are as follows:
\begin{enumerate}
	\item \citet{Chandrasekhar:PRSA1951}: The invariant theory of isotropic turbulence in magneto-hydrodynamics
	
	\item \citet{Chandrasekhar:PRSA1951_II}: The Invariant Theory of Isotropic Turbulence in Magneto-Hydrodynamics. II
	
	\item \citet{Chandrasekhar:PRSA1955_I}: Hydromagnetic turbulence. I. A deductive theory
	
	\item \citet{Chandrasekhar:PRSA1955_II}: Hydromagnetic turbulence II. An elementary theory
\end{enumerate}
  The first three papers  are in real  space, and they are generalization of the hydrodynamics equations of \citet{Karman:PRSA1938} and \citet{Kolmogorov:DANS1941Dissipation,Kolmogorov:DANS1941Structure} to MHD turbulence. 
  The fourth paper  attempts to employ Heisenberg's theory of turbulence to MHD turbulence (in spectral space). In the following discussion, we briefly sketch the results of these papers.

\subsection{Summary of the results of \citet{Chandrasekhar:PRSA1951,Chandrasekhar:PRSA1951_II} and \citet{Chandrasekhar:PRSA1955_I}}

 For isotropic and homogeneous MHD turbulence, \citet{Chandrasekhar:PRSA1951} derived equations for the second-order correlations of the velocity  and magnetic fields.  The derivation here is along the lines followed by \citet{Karman:PRSA1938}. Note that the equations for MHD turbulence are much more complex due to  more number of fields and  nonlinear terms than in HD turbulence.  As in other papers, Chandrasehkar follows rigorous and formal approach in these papers.   We skip the details due to their lengths and complexity, and provide only the leading equations of the papers.

The second-order correlation functions for the velocity and magnetic fields are given below:
 \bea
 \la u_i u_j' \ra = \frac{Q'}{l} l_i l_j - (l Q' + 2Q) \delta_{ij} , 
 \label{eq:Q} \\
  \la b_i b_j' \ra = \frac{H'}{l} l_i l_j - (l H' + 2H) \delta_{ij} . \label{eq:H} 
 \eea	
Here, \textbf{b} is the magnetic field, and $ u_j, b'_j $ represent the $ j $th components of the velocity and magnetic fields at the locations \textbf{r} and \textbf{r+l} respectively. Throughout the paper, the magnetic field is in  velocity units, which is obtained by dividing \textbf{b} in CGS unit with $ \sqrt{4\pi\rho} $, where $ \rho $ is the density of the fluid.  Note that the above correlation functions satisfy the incompressibility relations, $\partial'_j \la u_i u_j' \ra = 0$ and $ \partial'_j  \la b_i b_j' \ra =0$.

 As a sample, we present one of the  equations  derived by~\cite{Chandrasekhar:PRSA1951_II}:
 \bea
 \frac{\partial}{\partial l_m}  \la  (b_i u_m - b_m u_i) b_j' \ra
 & = &  \frac{\partial}{\partial l_m}   P(l_i \delta_{jm} - l_m \delta_{ij})   \nonumber \\
 & = & \frac{P'}{r} l_i l_j - (lP'+2P) \delta_{ij}  
 \eea
where $P$ is a scalar function, similar to $Q$ and $H$ of Eqs.~(\ref{eq:Q}, \ref{eq:H}).  Using the above equations and others, one of the inertial-range relations derived by Chandrasekhar is  
\bea
\la (u_1^2 + 2 b_2^2) u_1' \ra = -\frac{2}{15} \epsilon r,
\label{eq:Chandra_real1}
\eea
where $ \epsilon $ is the total dissipation rate, and $ u_1,b_1 $ are the longitudinal components along $ \hat{\bf l} $, while $ u_2, b_2 $ are components perpendicular to $ \hat{\bf l}$. {The above equation is a generalization of K41 relation to MHD turbulence.}

{For hydrodynamic turbulence, \citet{Loitsiansky} derived the following relations:
\bea
\int_0^\infty Q(l) l^4 dl = \mathrm{const.}
\eea
where $ Q(l) $ is the correlation function defined in Eq.~(\ref{eq:Q}). Using the dynamical equations of MHD, Chandrasehkar showed that Loitsiansky's integral remains constant to MHD turbulence as well.} In the second paper (\cite{Chandrasekhar:PRSA1951_II}), Chandrasekhar derived relations for the third-order correlation functions $ \la p u_i' u_j' \ra$ and $ \la p b_i' b_j' \ra$, where $ p $ is the pressure field. 

In  \citet{Chandrasekhar:PRSA1955_I}, Chandrasekhar derived a pair of differential equations  for the velocity and magnetic fields  at two different points and at two different times in terms of scalars. The derivation is quite mathematical and detailed, and is being skipped here.

\subsection{Summary of the results of \citet{Chandrasekhar:PRSA1955_II}}
 \citet{Chandrasekhar:PRSA1955_II}  generalized Heisenberg's theory for hydrodynamic turbulence to MHD turbulence. In this paper, the equations are in spectral space. One of the leading equations of the paper is 
\bea
&& -\frac{\partial}{\partial t}  \int_0^k dk' [E_u(k',t)+E_b(k',t) ] \nonumber \\
& = & 2 [ \nu \int_0^k k'^2 E_u(k') dk'+\eta \int_0^k k'^2 E_b(k') dk' ] \nonumber \\
&&+ \kappa \int_k^\infty  \left[ \sqrt{\frac{E_u(k')}{k'^3}} 
+ \sqrt{\frac{E_b(k')}{k'^3}}  \right] \times \nonumber \\ 
&& \int_0^k k'^2  [E_u(k') + E_b(k') ]dk', 
\label{eq:Chandra_Heisenberg}
\eea
where  $ E_u(k), E_b(k) $ are the energy spectra of the velocity and magnetic fields respectively, $ \eta $ is the magnetic diffusivity,  and $ \kappa $ is a constant. {Physical interpretation of Eq.~(\ref{eq:Chandra_Heisenberg}) is as follows. Without an external force, the energy lost by all the modes of a wavenumber sphere of radius $k$ is by (a) viscous and Joule dissipation in the sphere (the first term in the right-hand-side of Eq.(\ref{eq:Chandra_Heisenberg})), and (b) the nonlinear energy transfer from the modes inside the sphere to the modes outside the sphere (the second term in the right-hand-side of Eq.(\ref{eq:Chandra_Heisenberg})). The latter term is the total energy flux~(\cite{Verma:PR2004,Verma:book:ET}).}

Using the above equation, Chandrasekhar derived several results for the asymptotic cases, e.g., $ \nu \rightarrow 0 $ and $ \eta \rightarrow 0 $. For example, Chandrasekhar observed that for small wavenumbers ($ k \rightarrow 0 $), the velocity and magnetic fields are nearly equipartitioned, and they exhibit Kolmogorov's energy spectrum ($ k^{-5/3} $ ). However, at large wavenumbers, the magnetic and kinetic energies are not equipartitioned.  Quoting from his paper, ``in the velocity mode (kinetic-energy dominated case), the ratio of the magnetic energy to the kinetic energy tends to zero among the smallest eddies present (i.e., as $ k \rightarrow \infty $), while in the magnetic mode (magnetic-energy dominated case), the same ratio tends to about 2.6 as $ k \rightarrow \infty$."

\citet{Chandrasekhar:PRSA1951,Chandrasekhar:PRSA1951_II} and \citet{Chandrasekhar:PRSA1955_I,Chandrasekhar:PRSA1955_II}   are the first set of papers  on MHD turbulence. However, after these pioneering works, Chandrasekhar left the field somewhat abruptly. \cite{Sreenivasan:ARFM2019} ponders over this question in his review article.

A decade later, \citet{Kraichnan:PF1965MHD} and \citet{Iroshnikov:SA1964} brought next breakthroughs  in MHD turbulence. Thus, Chandrasekhar  pioneered the field of MHD turbulence.  We find that Chandrasekhar's results have  not been tested rigorously using numerical simulations and solar wind observations, and they have received less attention than   his other papers.  In the following discussion, we will briefly discuss some of the important papers after Chandrasekhar's work on MHD turbulence.

\section{Works in MHD turbulence between 1965 and 1990}
\label{sec:1965_1990}

\subsection{The energy spectrum $ k^{-3/2} $: Kraichnan and Iroshnikov}
In the presence of a mean magnetic field ($ {\bf B}_0 $), MHD has two kinds of Alfv\'{e}n waves that travel parallel and antiparallel to the mean magnetic field. \citet{Kraichnan:PF1965MHD} and \citet{Iroshnikov:SA1964} exploited this observation and argued that the Alfv\'{e}n time scale is the relevant time scale for  MHD turbulence.  Consequently, the interaction time for an Alfv\'{e}n wave of wavenumber $ k $ is proportional to $ (kB_0)^{-1} $.  Note that  the magnetic field including ${\bf B}_0 $ is in  velocity units.

Using these inputs and dimensional analysis, \citet{Kraichnan:PF1965MHD} and \citet{Iroshnikov:SA1964}  argued that the kinetic and magnetic energies are equipartitioned, and that the magnetic energy spectrum is
\bea
E_b(k) = A (\epsilon B_0)^{1/2} k^{-3/2},
\label{eq:Kraichan_3by2}
\eea
where $ A $ is a dimensionless constant. The above phenomenology predicts $ k^{-3/2} $ energy spectrum that differs  from Kolmogorov's $ k^{-5/3} $ spectrum, for which the relevant time scale is $ (k u_k)^{-1} $.  Note however that the solar wind turbulence tends to exhibit $ k^{-5/3} $ spectrum~[e.g., \cite{Matthaeus:JGR1982rugged}], however some authors report $ k^{-3/2} $ spectrum for the~solar~wind.

\subsection{Generalization by \cite{Dobrowolny:PRL1980}}
The MHD equations can be  written  in terms of Els\"{a}sser variables $ \bf{z}^\pm ={ \bf u} \pm {\bf b}$.  These variables represent the amplitudes of the  Alfv\'{e}n waves travelling in the opposite direction.  The nonlinear interactions between the Alfv\'{e}n waves yield  energy cascades. The fluxes of $ \bf{z}^+ $  and $ \bf{z}^- $  are $ \epsilon_{z^+} $ and $ \epsilon_{z^-} $ respectively, which are also  their respective dissipation rates.

 \citet{Dobrowolny:PRL1980} modelled the random scattering of Alfv\'{e}n waves.  They showed that the two fluxes are equal irrespective of the ratio $ z^+/z^- $, i.e.,
\bea
\epsilon_{z^+} = \epsilon_{z^-}.
\label{eq:equal_flux}
\eea
 \citet{Dobrowolny:PRL1980}  used these observations to explain depletion of cross helicity in the solar wind as it moves away from the Sun. Also, they derived $ k^{-3/2} $ energy spectrum for $ {\bf z}^\pm $, as in Eq.~(\ref{eq:Kraichan_3by2}).

\subsection{Field-theoretic calculation }

\citet{Fournier:JPA1982} employed field-theoretic methods to derive energy spectra $ E_u(k) $ and $ E_b(k) $, and the cross helicity spectrum $ H_c(k) $. They  employed the renormalization group procedure of  \citet{Yakhot:JSC1986}. The authors attempted to compute the renormalized viscosity and magnetic diffusivity, as well as vortex corrections. However, they were short of closure due to the complex nonlinear couplings of MHD turbulence.  There are  more field-theoretic works before 1990, but I am not describing them here due to lack of space.

Kraichnan and Iroshnikov's models dominated till 1990.   During this period, numerical simulations tended to support the $ k^{-3/2} $ spectrum [e.g., see \cite{Biskamp:PFB1989}], but they were not conclusive due to lower resolutions.  On the contrary, several solar wind observations [e.g., \cite{Matthaeus:JGR1982rugged}]  supported Kolmogorov's spectrum. In 1990's, new models and theories were constructed that support Kolmogorov's spectrum for MHD turbulence.  We describe these theories in the next section.

\section{Works between 1991 and 2010}
\label{sec:1991_2010}
As discussed earlier, \citet{Chandrasekhar:PRSA1955_II} argued that the kinetic and magnetic energy  follow $ k^{-5/3} $ spectrum as $ k \rightarrow 0 $.  More detailed works on Kolmogorov's spectrum for MHD turbulence followed after this work.

\subsection{Emergence of $ k^{-5/3} $ in MHD turbulence: \cite{Marsch:RMA1991}}

 \cite{Marsch:RMA1991}  considered a situation when the Alfv\'{e}nic fluctuations are much larger than the mean magnetic field.  In this case, the   nonlinear term ($ {\bf z}^\mp \cdot \nabla {\bf z}^\pm $)  dominates the linear term ($ {\bf B}_0 \cdot \nabla {\bf z}^\pm $). Here, usual dimensional arguments yields
\bea
\frac{E_{z^+}(k)}{E_{z^-}(k)} = \frac{K_+}{K_-} \left(  \frac{\epsilon_{z^+}}{\epsilon_{z^-}} \right)^2,
\label{eq:Marsch_5by3}
\eea
where  $ K _\pm$ are dimensionless constants. Note that the inertial-range fluxes  $ \epsilon_{z^+} $  and  $ \epsilon_{z^-} $ are unequal, unlike the predictions of \citet{Dobrowolny:PRL1980}  (see Eq.~(\ref{eq:equal_flux})). The inequality increases with the increase of the ratio $ E_{z^+}(k)/E_{z^-}(k)$.

Interestingly, the formulation of \citet{Dobrowolny:PRL1980}  too yields $ k^{-5/3} $ spectrum when the Alfv\'{e}n time is replaced by nonlinear time scale $ (k u_k)^{-1} $ (\cite{Verma:PR2004}).   \citet{Matthaeus:PF1989} attempted to combine the $ k^{-3/2} $ and $ k^{-5/3} $ models by proposing the  harmonic mean of the Alfv\'{e}n time scale and the nonlinear time scale as the relevant time scale. In their framework,  $ E(k) \sim k^{-5/3} $ for small wavenumbers, and $ E(k) \sim k^{-3/2} $ for larger wavenumbers. It turns out that the predictions of  \citet{Matthaeus:PF1989} are counter to weak turbulence theories where $ E(k) \sim k^{-3/2} $  should be active at small wavenumbers.

\subsection{Energy fluxes: Verma et al. [1994, 1996]}
For my Ph. D. thesis (\cite{Verma:thesis}), I wanted to verify which of the two spectra, $ k^{-5/3} $ and $ k^{-3/2} $, is valid for MHD turbulence.  We simulated several two-dimensiona (2D) MHD flows on $ 512^2 $ grids, and a single 3D flow on $ 128^3 $ grid. These runs had different  $ B_0 $ and $ z^+/z^- $.  We observed that the energy fluxes $ \epsilon_{z^\pm} $ satisfy Eq.~(\ref{eq:Marsch_5by3}) even when $ B_0 $ is five times larger than the fluctuations, and that the fluxes deviate significantly from Eq.~(\ref{eq:equal_flux}). Based on these observations, we concluded that Kolmogorov's model is more suited for MHD turbulence than Iroshnikov-Kraichnan model~(\cite{Verma:thesis,Verma:JGR1996DNS}).

\subsection{\citet{Politano:PRE1998} on structure functions}
Following similar approach as K41, \citet{Politano:PRE1998} showed that for MHD turbulence, the third-order structure function follows
\bea
\la ({\bf z'^{\pm} - z^\pm})^2  [({\bf z'^{\mp}} -{\bf z^\mp})\cdot \hat{l}] \ra = -\frac{4}{3} \epsilon_{z^\pm} l.
\eea
The above equations have a simple form because of the absence of  cross transfer between $ {\bf z}^+ $ and $ {\bf z}^- $. Note that the energy fluxes $ \Pi_{z^\pm} $ are constant in the inertial range~(\citet{Verma:book:ET}). The above relations translate to Kolmogorov's spectrum in Fourier space.

\citet{Politano:PRE1998} also derived  the third-order structure functions for the velocity and magnetic fields. These relations are more complex due to the coupling between the velocity and magnetic fields. Also refer to the complex relations in \cite{Chandrasekhar:PRSA1951}, which differ from  those of \citet{Politano:PRE1998}.

\subsection{Anisotropic MHD turbulence}
{Kolmogorov's $ k^{-5/3}$ theory and Iroshnikov-Kraichnan's $ k^{-3/2} $ theory assume the flow to be isotropic. However, this is not the case in MHD turbulence when a mean magnetic field is present. There are several interesting results for this case, which are discussed below.
	
\subsubsection{ \cite{Goldreich:ApJ1995}:}
 For anisotropic MHD turbulence, \cite{Goldreich:ApJ1995} argued that a critical balance is established between the Alfv\'{e}n time scale and nonlinear time scale, that is, $ k_\parallel B_0 \approx k_\perp z^\pm_{k_\perp} $.
 Using this assumption, Goldreich and Sridhar (1995) derived that 
 \bea
  E(k_\perp) = \epsilon^{2/3}  k_\perp^{-5/3},
 \eea
  which is  Kolmogorov's spectrum.
  
\subsubsection{Weak turbulence formalism:}
  For MHD turbulence with strong $ {\bf B}_0 $,
  \cite{Galtier:JPP2000} constructed a weak turbulence theory  and obtained 
  \bea
  \epsilon \sim \frac{1}{k_\parallel B_0} E_{z^+}(k_\perp)E_{z^-}(k_\perp) k_\perp^4.
  \label{eq:Weak_turb}
  \eea
When $ {\bf z}^+ $ and  $ {\bf z}^- $ have the same energy spectra,  Eq.~(\ref{eq:Weak_turb}) reduces to
    \bea
    E(k_\perp, k_\parallel) \sim  B_0^{1/2}  k_\perp^{1/2} k_\perp^{-2}.
  \eea
 Several numerical simulations support this prediction.  Note however that the solar wind turbulence exhibits nearly $ k^{-5/3} $ energy spectrum even though its  fluctuations  are five times weaker than the Parker field. This aspect needs a careful look.

\subsubsection{Anistropic energy spectrum and fluxes:}
In the presence of strong $ {\bf B}_0 $, the energy spectrum and energy transfers become anisotropic. \citet{Teaca:PRE2009} quantified the angular dependence of energy spectrum using {\em ring spectrum}. They showed that for strong $ {\bf B}_0 $, the energy tends to concentrate near the equator, which is  the region perpendicular to $ {\bf B}_0 $.   \citet{Teaca:PRE2009}  and \citet{Sundar:PP2017} also studied the anisotropic energy transfers using ring-to-ring energy transfers. 
In addition, \citet{Sundar:PP2017} showed that   strong magnetic field yields an inverse cascade of kinetic energy which may invalidate some of the assumptions made in  \citet{Goldreich:ApJ1995} and in \citet{Galtier:JPP2000}.
}

\subsection{Mean magnetic field renormalization}
Given that several solar wind observations, numerical simulations, and the works of \citet{Politano:PRE1998} support $ k^{-5/3} $ spectrum, it is quite puzzling what is going wrong with Kraichnan and Iroshnikov's  arguments on the scattering of Alfv\'{e}n waves. This led me to think about the effects of magnetic fluctuations on the propagation of  Alfv\'{e}n wave. 

In the presence of a mean magnetic field, MHD equations are nearly linear at large length scales. Alfv\'{e}n waves are the basic modes of the linearlized MHD equations. However,  the nonlinear term becomes significant at the intermediate and small scales (large wavenumbers).  Using renormalization group (RG) procedure, I could show that the an Alfv\'{e}n wave with wavenumber \textbf{k} is affected by an \textit{``effective" mean magnetic field}, which is  the  \textit{renormalized mean magnetic  field} (\citet{Verma:PP1999,Verma:PR2004}):
\bea
B_0(k) = C \epsilon^{1/3} k^{-1/3},
\label{eq:RGB0}
\eea
where $ C $ is a constant. Hence, an Alfv\'{e}n wave is not only affected by the mean magnetic field, but also by the waves with wavenumber near \textit{k}; this feature is called \textit{local interaction}. See Figure 2 for an illustration.  Note that \cite{Kraichnan:PF1965MHD} and \cite{Iroshnikov:SA1964} considered time scales based only on the mean magnetic field.
\begin{figure}
\centering\includegraphics[height=.15\textheight]{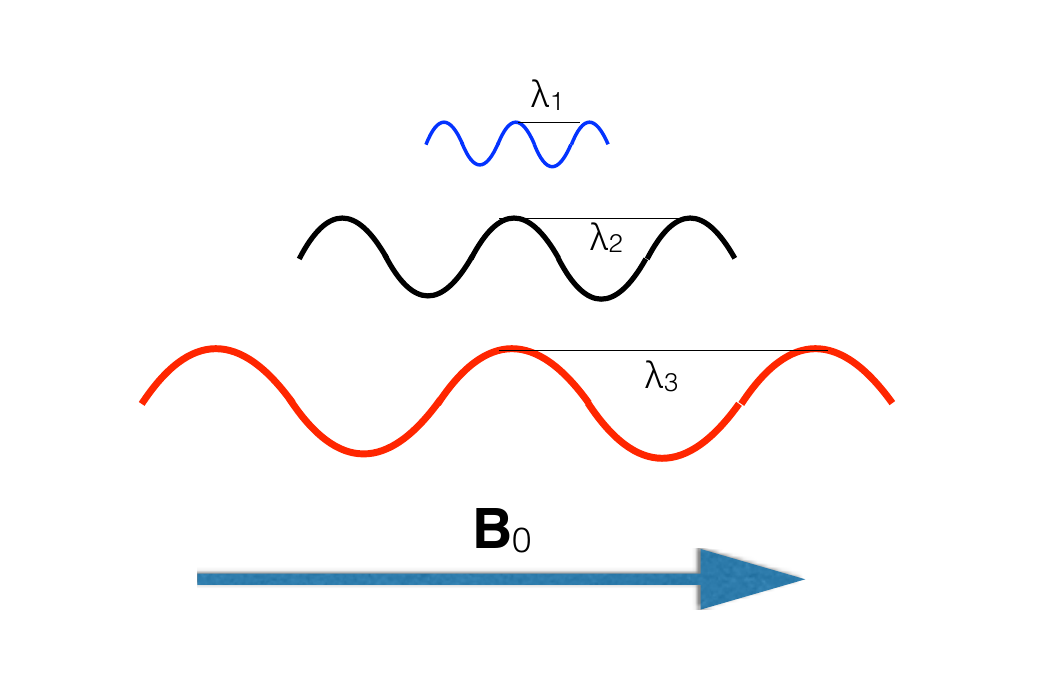}
\caption{A schematic diagram of multiscale Alfv\'{e}n waves. A fluctuation of wavelength $ \lambda_1 $ is affected by the  ``effective" or renormalized magnetic field $ B_0(k=1/\lambda_1) $ that scales as $ k^{-1/3} $.
Reproduced with permission from \cite{Verma:book:ET}. }
\end{figure}

Substitution of   $B_0(k) $ of Eq.~(\ref{eq:RGB0}) in Eq.~(\ref{eq:Kraichan_3by2}) yields 
\bea
E_u(k) \approx [\epsilon B_0(k)]^{1/2} k^{-3/2} \approx \epsilon^{2/3} k^{-5/3}.
\eea
Thus, we recover Kolmogorov's spectrum  in the framework of Kraichnan and Iroshnikov. Hence, there is consistency among various models. This argument is complimentary to those of \citet{Goldreich:ApJ1995}.

In the RG procedure of \citet{Verma:PP1999}, I went from large scales to small scales because the nonlinear interaction in MHD turbulence is weak at large scales. This is akin to quantum electrodynamics (QED) where particles (consider electrons) are free when they are separated by large distances.

\subsection{Renormalization of viscosity and magnetic diffusivity}

In the usual RG procedure of turbulence, we coarse grain the small-scale fluctuations (\cite{Yakhot:JSC1986}, \citet{McComb:book:Turbulence}). That is,  we average the small-scale fluctuations and  go to larger scales.   At small scales, the linearized MHD equations have viscous  and magnetic-diffusive terms. As we go to larger scales, the nonlinear terms enhance  diffusion, which is referred to as \textit{turbulent diffusion}. The effective diffusive constants in MHD turbulence are the  \textit{renormalized kinematic viscosity} and \textit{renormalized magnetic diffusivity}. 

 In \citet{Verma:PRE2001}, \citet{Verma:Pramana2003Nonhelical}, \citet{Verma:Pramana2003Helical}, and \citet{Verma:PR2004}, I  implemented the above scheme using the self-consistent procedure of \cite{McComb:book:Turbulence,McComb:book:HIT}, and computed the renormalized   viscosity and magnetic diffusivity.  This self-consistent procedure was useful in circumventing the difficulties faced by \citet{Fournier:JPA1982} and others. Compared to the procedure of \cite{Yakhot:JSC1986}, McComb's scheme has less parameters to renormalize.  For tractability, I focussed on the following two limiting cases:

\subsubsection{Cross helicity $ H_c = 0 $:} This assumption leads to major simplification of the calculation. I could show that
\bea
\nu(k) & = & \sqrt{K} \nu_* \epsilon^{1/3} k^{_4/3}, \\
\eta(k) & = & \sqrt{K}\eta_* \epsilon^{1/3} k^{-4/3}, \\
E(k) &= &  K \epsilon^{2/3} k^{-5/3}
\eea
are consistent solutions of RG equations. Thus, we show that the kinetic and magnetic energies  exhibit $ k^{-5/3} $ energy spectra. 

\subsubsection{Non-Alfv\'{e}nic case, $ z^+ \gg z^- $:} This limiting case corresponds to large cross helicity.  Again, a self-consistent RG procedure yields $ k^{-5/3} $ spectrum for the Els\"{a}sser variables.

\subsection{\citet{Boldyrev:PRL2006} revives $ k^{-3/2} $ spectrum}

\citet{Boldyrev:PRL2006} hypothesized that  the inertial-range fluctuations of MHD turbulence have certain dynamical alignments that yields interaction time scale as 
\bea
T_k \sim (k u_k \sin \theta_k)^{-1} \sim   (k u_k \theta_k)^{-1},
\eea
 where $ \theta_k $ is the angle between the velocity and magnetic fluctuations at  the scale of $ k^{-1} $.  \citet{Boldyrev:PRL2006} argued that $ \theta_k \sim k^{-1/4} $. Using dimensional analysis, we obtain
 \bea
 \theta_k \sim k^{-1/4} (\epsilon/B_0^3)
^{1/4} ,
\eea
 substitution of which in the flux equation yields
 \bea
 \Pi \sim \epsilon \sim \frac{u_k^2}{T_k} \sim k u_k^3 k^{-1/4} (\epsilon/B_0^3)^{1/4}.
\eea
The above equation was inverted to obtain the following energy spectrum:
\bea
E_u(k) \sim (\epsilon B_0)^{1/2} k^{-3/2},
\label{eq:Boldyrev}
\eea
which  is same as that predicted by \cite{Kraichnan:PF1965MHD} and \cite{Iroshnikov:SA1964}. Boldyrev and coworkers performed numerical simulations and observed consistency with  the above  predictions.  Thus, $ k^{-3/2} $ spectrum has come back with vengeance.

\subsection{Energy fluxes of MHD turbulence}
MHD turbulence has six energy fluxes, in contrast to single flux of hydrodynamic turbulence~(\cite{Dar:PD2001,Verma:PR2004,Debliquy:PP2005}). The energy fluxes from the velocity field to the magnetic field are responsible for  dynamo action, or amplification of magnetic field in astrophysical objects~(\cite{Brandenburg:PR2005,Kumar:JOT2015,Verma:JOT2016}). Energy fluxes can also help us decipher the physics of MHD turbulence, e.g.,  in \citet{Verma:JGR1996DNS}.  We cannot describe details of energy flux in this short paper; we refer the reader to \cite{Verma:PR2004,Brandenburg:PR2005}; and \cite{Verma:book:ET} for details.

{
	\section{Possible approaches to reach the final theory of MHD turbulence}

As discussed above, we are far from the final theory of MHD turbulence.  Future high-resolution simulations and data from space missions may help resolve this long-standing problem. I believe that the following explorations would provide important clues for MHD turbulence:
\begin{enumerate}
	\item Measurements of the time series of  the inertial-range Alfv\'{e}n waves would help us explore the wavenumber dependence of  $ B_0 $~(\cite{Verma:PP1999}).
	
	\item The energy fluxes of $ {\bf z^\pm} $, $ \epsilon_{z^\pm} $, are approximately equal in the Iroshnikov-Kraichnan phenomenology, but not so in Kolmgoorov-like phenomenology for MHD turbulence. \citet{Verma:JGR1996DNS} showed that $ \epsilon_{z^\pm} $  for 2D MHD turbulence follow Kolmogorov-like theory. But, we need to extend this study to three dimensions and for high resolutions. The findings through these studies will also help estimate the turbulent heating in the solar wind and in the solar corona.
	
	\item Recent spacecrafts are providing high-resolution solar wind and corona data, which can be used for investigating MHD turbulence. These studies would compliment numerical studies.
\end{enumerate}

We hope that above studies would be carried out in near future, and we will have a definitive theory of MHD turbulence soon.
 }

\section{Summary}
\label{sec:summary}
In this paper, I surveyed the journey of MHD turbulence, starting from the pioneering works of Chandrasekhar.  Chandrasekhar attempted to model the structure functions and energy spectra of MHD turbulence. Unfortunately, Chandrasekhar's papers on hydrodynamic and hydromagnetic turbulence did not attract significant attention in the community.   \citet{Sreenivasan:ARFM2019}, who studied this issue in detail, points out the following possible reasons for the above. Chandrasekhar's papers are typically more mathematical than a typical paper on turbulence.  As written in \citet{Sreenivasan:ARFM2019}, ``what mattered to Chandra was what the equations revealed; everything else was superstition and complacency."   Thus, Chandrasekhar did not make significant effort to extract physics from mathematical equations, unlike the other stalwarts of the field (e.g., Batchelor, Taylor, Kolmogorov).  

\citet{Sreenivasan:ARFM2019} points out another  factor  that drifted Chadrasekhar from the turbulence community. Chandrasekhar sent one of his important manuscripts on turbulence to the  Proceedings of  Royal  Society, but the paper was rejected. This paper was eventually published in Physical Review (\cite{Chandrasekhar:PR1956}), but it  contained several incorrect assumptions~(\cite{Sreenivasan:ARFM2019}). When these assumptions were criticised by Kraichnan and others, Chandrasekhar did not take them kindly and  left the field of turbulence abruptly. Refer to \citet{Sreenivasan:ARFM2019} for details on this topic.

More work on MHD turbulence followed 10 years after Chandrasekhar left this field.  I divided these works in two temporal regimes: between 1965 to 1990, and between 1991 to 2010. The first period was dominated by Kraichnan and Iroshnikov's $ k^{-3/2} $ model, which is based on the scattering of Alfv\'{e}n waves.  Till 1990, the community appears to believe in the validity of this theory, even though several astrophysical observations supported $ k^{-5/3} $ spectrum.  From 1991 onwards, there were a flurry of models and calculations that support Kolmogorov-like spectrum ($ k^{-5/3} $) for MHD turbulence. However, in 2006, Boldyrev and coworkers argued in favour of $ k^{-3/2} $ spectrum.  Hence, the jury is not yet out.  More detailed diagnostics have to be performed to arrive at the final theory of MHD turbulence. 

At present, there is a lull in this fields.  We hope that in near future, we will be able to completely understand the underlying physics of MHD turbulence, a journey that started with Chandrasekhar's pioneering work.

\section*{Acknowledgements}
I enjoyed participating in the conference ``Chandra's Contribution in Plasma Astrophysics". I thank the organizers, especially Ram Prasad Prajapati, for the invitation. {I am grateful to Katepalli Sreenivasan (Sreeni) for insightful discussions on the contributions and  work style of Chandrasekhar. In fact, the present paper is inspired by Sreeni's ARFM article, {\em Chandra's Fluid Dynamics}. I also thank Sreeni for numerous useful suggestions on this paper. }In addition, I thank all my collaborators---Melvyn Goldstein, Aaron Roberts, Gaurav Dar, Rodion Stepanov, Franck Plunian,  Daniele Carati, Olivier Debliquy, Riddhi Bandyopadhyay, Stephan Fauve, and Vinayak Eswaran---for wonderful discussions on MHD turbulence, and to Anurag Gupta for useful comments. 
\vspace{-1em}



\end{document}